# Coherent population trapping combined with cycling transitions for quantum dot hole spins using triplet trion states

Samuel G. Carter,[1] Stefan C. Badescu,[2] Allan S. Bracker,[1] Michael K. Yakes,[1†] Kha X. Tran,[3] Joel Q. Grim,[1] and Daniel Gammon[1]

[1] Naval Research Laboratory, 4555 Overlook Ave. SW, Washington, DC 20375, USA

[2] Air Force Research Laboratory, Sensors Directorate, Wright Patterson AFB, Ohio 45433, USA

[3] NRC Research Associate at the Naval Research Laboratory, 4555 Overlook Ave. SW, Washington, DC 20375, USA

Optical spin rotations and cycling transitions for measurement are normally incompatible in quantum dots, presenting a fundamental problem for quantum information applications. Here we show that for a hole spin this problem can be addressed using a trion with one hole in an excited orbital, where strong spin-orbit interaction tilts the spin. Then, a particular trion triplet forms a double Λ system, even in a Faraday magnetic field, which we use to demonstrate fast hole spin initialization and coherent population trapping. The lowest trion transitions still strongly preserve spin, thus combining fast optical spin control with cycling transitions for spin readout.

One of the most attractive features of self-assembled quantum dots (QDs) is their strong coupling to light, which makes them excellent photon emitters and also allows for fast optical manipulation of spin states. The ideal energy level system for optical control of the ground state spin is a Λ system in which both spin states couple to one excited state. The double-Λ system shown in Fig. 1(b) is commonly achieved for singly-charged QDs by applying a magnetic field perpendicular to the optical axis (Voigt geometry). This charged exciton (trion) system has been used to demonstrate optical spin initialization [1], ultrafast spin rotations [2,3], and coherent population trapping (CPT) [4–7]. For spin readout, however, a cycling transition is desired that results in the emission of many photons without changing the spin state. This has been accomplished for singly-charged QDs by applying a magnetic field parallel to the optical axis (Faraday geometry), giving a double two-level system shown in Fig. 1(a), with a demonstration of single-shot spin readout [8].

*sam.carter@nrl.navy.mil
†Present address: Air Force Office of Scientific Research, Arlington, VA 22203

Achieving both types of energy level systems for QDs in one geometry has been quite challenging, as the requirements for spin control and readout are conflicting. Some progress has been made with QDs in photonic crystal cavities and waveguides that enhance one set of transitions polarized with the cavity and inhibit the other set, making one set useful for initialization and the other for readout [9–12]. This technique requires a strong Purcell effect and alignment of the transition polarization with the cavity. Other approaches have been to use a spin selective AC Stark shift to optically change the energy level structure [13] or to make use of light-hole excitons [14–16]. There have also been efforts with pairs of coupled QDs, which have additional degrees of freedom that allow both Λ systems for control and cycling transitions for readout [17,18]. This system, however, is more difficult to produce.

In this letter we report a new, simple approach to this challenge for a hole spin in a single QD using trion states in which one hole is in an excited orbital (often called "hot" trions). In a Faraday magnetic field, the lowest trion states form the double two-level system in Fig. 1(a), which has two cycling transitions, as is well known [8,19]. The next optical transitions involve exciting an electron in the lowest orbital ($e_0$) and a hole in the 1$^{st}$ excited orbital ($h_1$). Using photoluminescence excitation (PLE) spectroscopy, we find that these triplet transitions can be very sharp (~8-10 μeV), and one particular triplet forms a double Λ system in a Faraday magnetic field [see Fig. 1(b)]. This happens because transitions that normally would be forbidden by spin selection rules become allowed due to spin orbit coupling for a hole in higher orbitals. We use this Λ system to demonstrate fast hole spin initialization as well as CPT. This system provides for both efficient readout and fast optical control and also has the advantage of improved coherence of the hole spin over an electron spin, due to a weaker hyperfine interaction [5,20–24].

The InGaAs QDs are grown by molecular beam epitaxy on an n-doped GaAs substrate within a distributed Bragg reflector planar cavity [25]. A vertical *n-i-p-i-p* diode within the cavity is used to charge the QD with a single hole. Experiments are performed between 2.8 and 5 K, at biases near the transition from one hole to two holes in order to vary the spin relaxation rate [30], either preventing or allowing optical pumping of the hole spin.

The states considered here and their fine structure are shown in Fig. 2(a). The ground state has one hole in the lowest orbital ($h_0$). The lowest trion state ($X^+_{h0h0}$) has two holes in $h_0$ and an electron in $e_0$. The hot trion state ($X^+_{h0h1}$) has one hole in $h_1$, another hole in $h_0$, and the electron in $e_0$. The $e_0$, $h_0$, and $h_1$, spins are represented by ↑, ⇑$^0$, and ⇑$^1$, respectively. The holes should be primarily heavy holes with $m_j = \pm 3/2$. For $X^+_{h0h0}$ the holes must form a spin singlet, giving two states $S^0_{\pm 1/2}$ with different electron spin projections. The fine structure of hot trion states has been explained in previous studies [31–37]. For $X^+_{h0h}$ the holes can be in a singlet state or triplet states split by the isotropic hole-hole exchange energy $\Delta_{hh}$. These trion states are labeled by $S^1_{m_f}$ and $T^1_{m_f}$, where $m_f = m_s + m_J$ is the total spin projection for the electron ($m_s$) and two holes ($m_J$). As has been reported previously, anisotropy in the hole-hole exchange interaction results in a shift $\delta_{hh}$ of the $m_J = 0$ triplet $T^1_{\pm 1/2}$ relative to $m_J = \pm 3$ triplets ($T^1_{\pm 5/2}$ and $T^1_{\pm 7/2}$), as well as shifts of the singlet $S^1_{\pm 1/2}$ [33]. The electron-hole exchange interaction further splits $T^1_{\pm 5/2}$ and $T^1_{\pm 7/2}$ by $2\Delta_{eh}$, which is much weaker than $\Delta_{hh}$. Asymmetric electron-hole exchange is neglected. Each state is doubly degenerate at zero magnetic field.

PLE spectroscopy is performed by tuning the laser through the $X^+_{h0h1}$ transitions and measuring emission from $X^+_{h0h0}$ with a CCD spectrometer. In Fig. 2(b), the emission spectrum of $X^+_{h0h0}$ at B=0 T shows a single main line. The weak, lower energy line comes from $X^{2+}$. Figure 2(c) plots the PLE spectrum of the next transitions of the QD, integrating emission from $S^0_{\pm 1/2}$, showing three sharp triplet lines ~3 meV below a broader singlet line. From the transition energies, we obtain $\Delta_{hh}$~3 meV, $\delta_{hh} = 0.605$ meV, and $\Delta_{eh}= 0.234$ meV. From the singlet-triplet structure, we infer that the two holes must be in different orbitals, and the small energy separation from $X^+_{h0h0}$ (13-17 meV compared to ~50 meV in Ref. [32] for $e_1 h_1$ transitions) indicates that the electron must be in $e_0$. The triplet transition linewidths are 8-10 μeV, and the singlet transition linewidth is 52 μeV, which are much sharper than in previous studies [32–34], in which hot trions were measured at higher energies above the lowest exciton (>20 meV). Higher energy PLE lines are also observed for this QD (not shown) that have similar singlet-triplet patterns but broader linewidths. These lines are assigned to higher hole states ($h_2$, $h_3$), and the broader linewidths are attributed to faster relaxation, probably related to proximity to the LO phonon energy. We note that none of these transitions should be allowed for p-like $h_1$ and s-like $e_0$, due to zero electron-hole overlap, indicating mixed symmetry of orbitals [38,39]. Also, the $T^1_{\pm 7/2}$ transitions are nominally forbidden by spin selection rules since they do not change the spin projection by $\Delta m_f = \pm 1$. These "dark" trions have been observed in previous studies as well [33,34].

More insight into these transitions is obtained by applying a magnetic field along the growth direction and optical axis (Faraday geometry). Figure 3(a) displays spectrally-resolved resonance fluorescence of $X^+_{h0h0}$ at B=1.5 T when driving the $\Uparrow^0 - S^0_{+1/2}$ and $\Downarrow^0 - S^0_{-1/2}$ transitions resonantly. Spectra are taken near the single hole stability edge for fast hole spin relaxation to prevent optical pumping. When driving one transition, there is no sign of emission from the other allowed transition or from forbidden cross transitions (e.g. $\Uparrow^0 - S^0_{-1/2}$), indicating these transitions preserve spin (cycling transitions).

In Fig. 3(b) the PLE of the triplet transitions is displayed at B=2 T for $\sigma^+$ and $\sigma^-$ excitation, integrating emission from both $S^0_{\pm 1/2}$ lines. Both the $T^1_{\pm 7/2}$ and $T^1_{\pm 5/2}$ transitions split into two lines with opposite circular polarizations while $T^1_{\pm 1/2}$ splits into four lines, with two lines $\sigma^+$ and two lines $\sigma^-$. The energy level diagram in Fig. 3(c) displays all of these transitions. Four transitions (solid lines) are expected from spin selection rules in which a spin $\pm 1$ $e_0 h_1$ exciton ($\downarrow\Uparrow^1$ or $\uparrow\Downarrow^1$) is generated with $\sigma^\pm$ polarization in addition to the resident $h_0$ hole. The four unexpected transitions (dashed lines) correspond to generating a spin $\pm 2$ $e_0 h_1$ exciton ($\uparrow\Uparrow^1$ or $\downarrow\Downarrow^1$) with $\sigma^\mp$ polarization. This provides a $\sigma^+$ $\Lambda$ system for $T^1_{-1/2}$ and a $\sigma^-$ $\Lambda$ system for $T^1_{+1/2}$. The double $\Lambda$ system also occurs for $S^1_{\pm 1/2}$ (not shown). Moreover, a similar transition pattern occurs for the next higher shell trion ($X^+_{h0h2}$). This pattern has been observed in all three QDs measured.

The unexpected transitions that give rise to the double $\Lambda$ system are allowed due to spin-orbit coupling that is strong for $h_1$ and weak for $h_0$. The spin-orbit coupling can be understood in terms of an effective magnetic field due to motion in the confinement potential [40]. More information is given in the Supplemental Material [25]. The in-plane component of the effective magnetic field is responsible for spin mixing combined with orbital mixing. The state of the predominantly $h_1$ spin up state can thus be written as $\Uparrow^1_t = \alpha |h_1\rangle|\Uparrow\rangle + \sum_{i>1} \beta_i |h_i\rangle|\Downarrow\rangle$, where $|h_i\rangle$ are the orbital states, $|\Uparrow\rangle$ and $|\Downarrow\rangle$ are pseudospin states, and $\alpha$ is expected to be nearly 1. This "tilted" spin $\Uparrow^1_t$ explains all of the unexpected transitions and their polarizations in Fig. 3(c). For example, the triplet state $(\uparrow\Uparrow^0\Downarrow^1 + \uparrow\Downarrow^0\Uparrow^1)/\sqrt{2}$ is given a small component $\uparrow\Downarrow^0\Downarrow^i$, with one hole in an excited orbital, by substituting $\Uparrow^1 \to \alpha \Uparrow^1 + \beta \Downarrow^i$ in the second term. This makes the transition from the ground state $\Downarrow^0$ allowed with $\sigma^-$ polarization, as observed. Other forbidden transitions that are not observed (e.g. $\Downarrow^0$ to $\downarrow\Uparrow^0\Uparrow^1$) should still be very weak since they require a spin mixing in $h_0$. We also note that even with $\beta_i \ll 1$, there can be a strong effect on the optical transitions due to changes in the electron-hole overlap. The overlap is weak for the nominally odd parity $h_1$ but can be much larger for even parity orbitals (e.g. $h_3$) that are mixed in, amplifying the effect of the spin mixing terms on optical transitions. From symmetry arguments and energy separations, we expect that the dominant mixing term is with $h_3$, a nominally even parity $d$ orbital with two nodes along the same axis as $h_1$ [25].

There is also an effective spin-orbit magnetic field component along the growth direction (z) that results in spin-dependent mixing of orbitals, $h_1$ and $h_2$ ($p_x$ and $p_y$), while preserving the spin projection along $z$ [41]. This does not tilt the spin and change spin-selection rules, but it does result in an anisotropic hole-hole exchange interaction [41,42]. This leads to a shift of $T^1_{\pm 5/2}$ and $T^1_{\pm 7/2}$ down in energy more than $T^1_{\pm 1/2}$, resulting in the relative anisotropic exchange shift ($\delta_{hh}$) observed in Fig. 2.

We have shown that the strong spin-orbit interactions of the excited hole in the hot trion gives the double $\Lambda$ system that has been missing in the Faraday geometry. Next we show that this double $\Lambda$ system can be used to control the ground state hole spin, starting with the ability to optically pump into a particular spin ground state. When driving a particular trion transition, optical spin pumping occurs when relaxation of the trion has some chance of returning the system to the opposite hole spin state being driven. This can happen with any of the transitions in Fig. 3(c), including the lowest energy singlet $S^0_{\pm 1/2}$, though slowly [8]. We obtain *fast* optical pumping using the new transitions in the trion triplet. In Fig. 4 we focus on the $\Lambda$ system formed with $T^1_{-1/2}$ and measure time-resolved optical pumping.

The hole spin state is first randomized by pulsed excitation for 30 ns with linearly polarized light at 1388 meV, where transitions are broad enough to excite both hole spin states. Then a 100 ns pulse excites one of the triplet transitions. Figure 4(a) displays the time-correlated-photon-counting of emission from $S^0_{-1/2}$ during this pulse for excitation of the $\Uparrow^0 - T^1_{-1/2}$ and $\Downarrow^0 - T^1_{-1/2}$ transitions. The emission turns on rapidly with the pulse, followed by an exponential decay to nearly zero as the hole spin is pumped. At a peak power of 1 µW, the decay time constants are 7.3 ns and 20.5 ns for the $\Uparrow^0 - T^1_{-1/2}$, and $\Downarrow^0 - T^1_{-1/2}$ transitions, with initialization fidelities of 99% and 97%, respectively [25]. The difference in peak intensities is attributed to the relative oscillator strengths. The differences in the decay times can be explained by the trion relaxation processes illustrated in Fig. 4(b), in which $T^1_{-1/2}$ primarily relaxes to $S^0_{-1/2}$, followed by emission to $\Downarrow^0$ (see Supplemental Material). Driving $\Uparrow^0 - T^1_{-1/2}$ is thus more likely to change the spin state than driving $\Downarrow^0 - T^1_{-1/2}$. As the excitation power is increased to saturation [see inset of Fig. 4(a)], the pumping time through the $\Uparrow^0 - T^1_{-1/2}$ transition is only 1.5 ns, comparable to s-shell trion pumping times in the Voigt geometry [1] and about 3 orders of magnitude shorter than s-shell pumping times in the Faraday geometry [8].

With a double Λ system in the $T^1_{\pm 1/2}$ states, coherent control of the hole spin now becomes possible in the Faraday geometry. In Fig. 5 we demonstrate CPT through the trion triplet. This experiment was done on a different QD from the same sample that shows very similar behavior to that measured in Figs. 2-4 [25]. As shown in Fig. 5(a-b), a pump laser is tuned to the $\Downarrow^0 - T^1_{-1/2}$ transition and a weaker sideband probe is tuned near the $\Uparrow^0 - T^1_{-1/2}$ transition, generated by an electro-optic phase modulator. The frequency shift of the sideband is determined directly by the microwave modulation frequency, eliminating frequency drift between pump and probe. Emission from $\Downarrow^0 - S^0_{-1/2}$ is plotted vs the probe detuning in Fig. 5(c) for a series of pump powers. The dip observed at 40.5 µeV corresponds to the formation of a "dark" state when the frequency difference is equal to the ground state hole spin splitting. This dark state consists of a superposition of $\Uparrow^0$ and $\Downarrow^0$ with coefficients determined by $\Omega_{\text{pump}}/\Omega_{\text{probe}}$, the ratio of Rabi frequencies [43]. The probe power, which varies some with the modulation frequency, is about 7% of the pump power, giving an estimate of $\Omega_{\text{probe}} = \Omega_{\text{pump}}/4$. To observe a strong CPT dip, $\Omega^2_{\text{pump}}/\Gamma \gg 1/T_2$, where $\Gamma$ is the excited state relaxation rate and $T_2$ is the spin coherence time. We numerically model the CPT data using a 3-level Λ system [25,44,45] that includes spectral wandering by weighted averages over variations in the spin splitting and excited state energy. The model calculations in Fig. 5(d), taken at the experimentally determined values of $\Omega_{\text{pump}}$ [25], fit the experimental data quite well at low powers. At higher powers, the dip appears to have relatively sharp edges, and the broad peak has red-shifted a little, which both likely come from nuclear polarization effects that are not captured in the model [46]. This model gives a $T_2^*$ of about 9 ns, which is short compared to similar measurements in the Voigt geometry with $T_2^* > 100$ ns [6,7]. To our knowledge no measurements of the hole spin $T_2^*$ for InAs QDs have been performed in the Faraday geometry, in which holes have a very different g-factor and the hyperfine interaction is stronger [7,20], so further study is needed to characterize and understand this behavior.

This Letter demonstrates that hot trion states provide additional energy levels for control and readout of spin. In particular the triplet $T^1_{\pm 1/2}$ states provide a double Λ system in a Faraday magnetic field, due to spin-forbidden transitions that are consistently allowed. These transitions arise from tilting of the excited hole spin by the spin-orbit interaction and occur for typical QD samples without using any special techniques. Using this system, we demonstrate fast initialization and CPT while also showing that the lowest trion transitions strongly preserve spin. This addresses the long-standing challenge in QDs of combining efficient spin readout with fast, coherent spin control in one geometry. These higher energy transitions also have the important advantage of being spectrally separated from the emission, eliminating laser scatter with spectral filtering.

This work was supported by the US Office of Naval Research, the Defense Threat Reduction Agency (Grant No. HDTRA1-15-1-0011), and the Air Force Office of Scientific Research (Award No. FA9550-AFOSR-17RYCOR500).

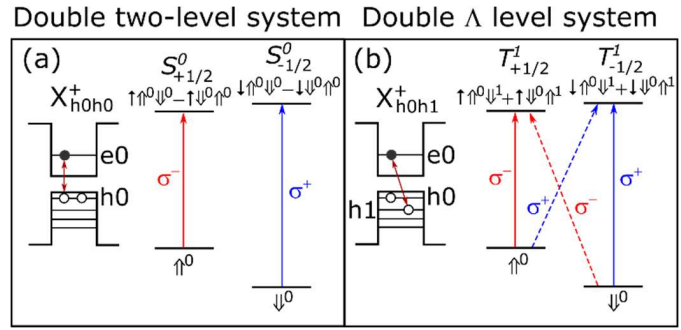

FIG. 1. (a) Double two-level system for the s-shell charged exciton in a Faraday magnetic field. (b) Double $\Lambda$ system for a "hot" trion with one hole in an excited orbital in a Faraday magnetic field. Electron (hole) spins in $e_0$ ($h_0$) are represented by $\uparrow$ ($\Uparrow^0$), and a hole spin in $h_1$ is represented by $\Uparrow^1$.

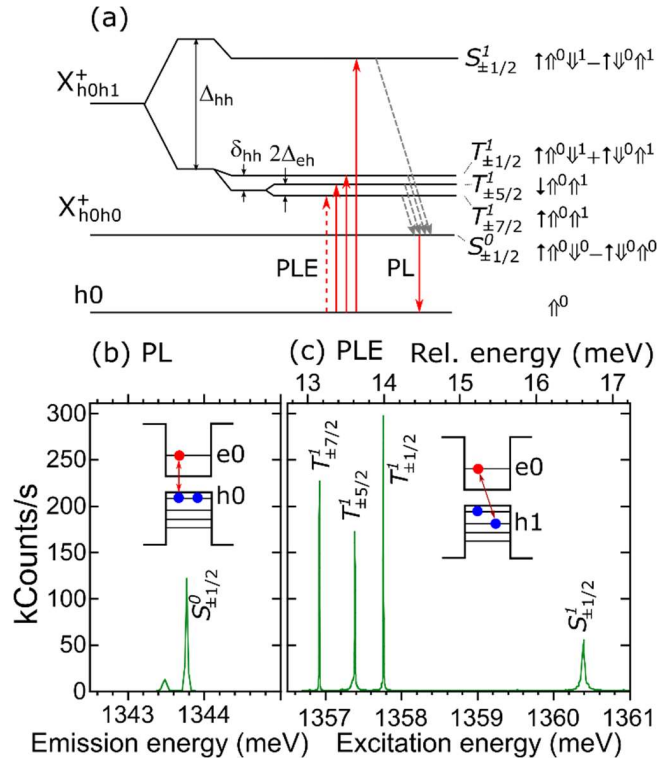

FIG. 2. (a) Energy level diagram of a QD charged with a hole, showing optical transitions to the singlet and triplet hot trions $X^+_{h0h1}$, non-radiative relaxation to the lowest energy trion $X^+_{h0h0}$, and emission to the ground state $h_0$. (b) PL emission spectrum at B=0 T for the lowest energy trion $X^+_{h0h0}$. (c) PLE spectrum of $X^+_{h0h1}$ at B=0 T, collecting emission from $X^+_{h0h0}$, with a laser power of 1 μW. Schematics of the QD electron and hole energy levels showing occupation is inset in (b) and (c).

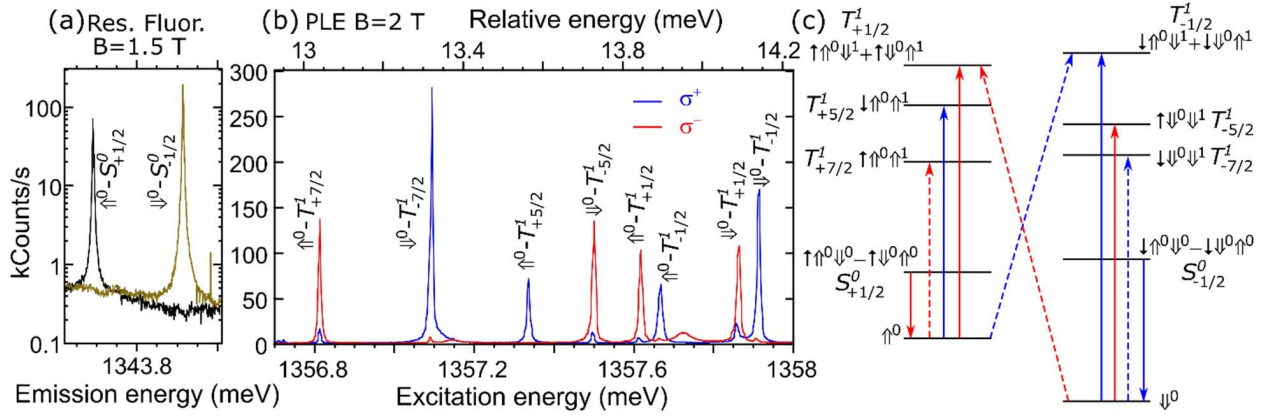

FIG. 3. (a) Spectrally resolved resonance fluorescence of $X^+_{h0h0}$ at B=1.5 T with the laser linearly polarized at a power of 10 nW. (b) PLE of the triplet transitions at B=2.0 T, collecting light from both emission lines of $X^+_{h0h0}$, for $\sigma^+$ and $\sigma^-$ excitation polarizations at 1 µW. The sample is biased at $V_{\text{bias}} = -0.95$ V at the charge stability edge for (a) and (b). (c) Energy level diagram showing the triplets and the lowest energy singlets, with red ($\sigma^-$) and blue ($\sigma^+$) arrows showing the expected (solid) and unexpected (dashed) optically allowed transitions.

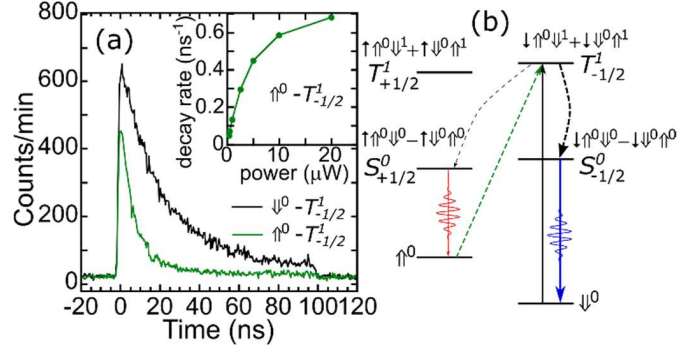

FIG. 4. (a) Time-correlated photon counting of a 100 ns pulse exciting two triplet transitions, with B=1.5 T and $V_{\text{bias}} = -0.97$ V. Emission is collected from $S^0_{-1/2}$. The inset plots the decay rate as a function of drive power for the $\Uparrow^0 - T^1_{-1/2}$ transition. (b) Energy level diagram showing excitation of $T^1_{-1/2}$ by either of two transitions, followed by non-radiative relaxation (curved dashed arrows), primarily to $S^0_{-1/2}$, and then emission into $\Downarrow^0$.

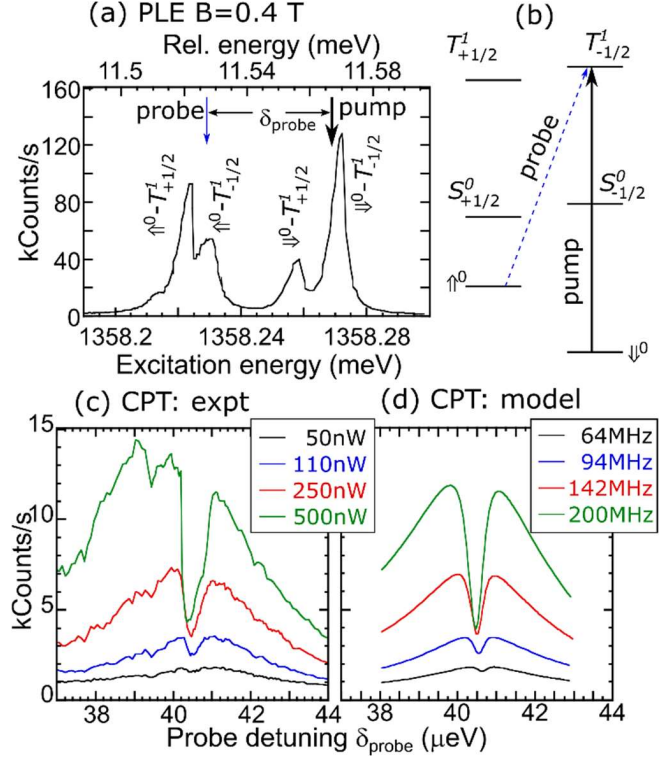

FIG. 5. (a) PLE spectrum of $T^1_{\pm 1/2}$ at B=0.4 T for the 2$^{nd}$ QD, using linearly polarized excitation at 0.6 μW. The bias is set to the stability edge for fast hole spin relaxation. (b) Energy level diagram showing the pump and probe driving the $T^1_{-1/2}$ Λ system. (c) CPT measurements for a series of pump powers at a bias 30 mV away from the stability edge, where hole spin relaxation is much slower. The probe power is 7% of the pump, and both pump and probe are polarized $\sigma^+$. The arrows in (a) indicate the positions of the pump and probe, correcting for the Stark shift between the two biases. (d) CPT model for a series of Rabi frequencies matching experiment.

# Supplemental Material

## I. Sample heterostructure

The sample is grown by molecular beam epitaxy on an n-doped GaAs substrate with the following structure:

Substrate
10 period Bragg mirror:
    82 nm Si doped AlAs    (n)
    69 nm Si doped GaAs
82 nm Si doped AlAs
30 nm Si doped GaAs    (n)
24 nm undoped GaAs    (i)
10 nm Be doped GaAs    (p)
75 nm undoped GaAs    (i)
InAs QD
25 nm undoped GaAs    (i)
114 nm Be doped GaAs    (p)
4 period Bragg mirror, Be doped:
    82 nm Be doped AlAs    (p)
    69 nm Be doped GaAs

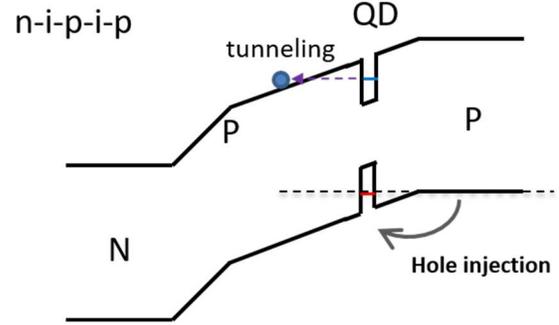

Fig. S1. Illustration of sample band structure, showing hole injection and electron tunneling out of the dot. Bragg mirrors are not shown.

In this n-i-p-i-p diode structure, the QD is placed nearest the p-doped region in order to inject a hole into the QDs. As illustrated in Fig. S1, an intermediate p-doped 10 nm layer in between the n-type region and the QD is included to reduce the built-in electric field, resulting in lower diode currents and thus reduced heating.

    In this device there is an issue with optically excited electrons tunneling out of the QD before recombination occurs. We observe this in PL bias maps, in which the intensity of a PL line gets weaker with reverse bias. As the reverse bias increases (more negative biases), the electric field increases, resulting in a thinner electron tunneling barrier, faster tunneling, and reduced PL intensity. This results in weaker emission, and it can also significantly disrupt optical cycling transitions. For example, excitation of the $\Uparrow^0 - S^0_{+1/2}$ transition, followed by the electron tunneling out, results in two holes in the singlet state. One hole will quickly tunnel out of the QD (at a bias where only one hole is stable), leaving a randomly oriented hole spin. This effect is seen to significantly reduce the spin flip time (optical pumping time) for the $S^0_{\pm 1/2}$ transitions, preventing single shot readout. The inclusion of an AlGaAs barrier between the QD and n-type region should prevent this tunneling problem.

## II. PLE Spectrum of 2$^{nd}$ QD

The PLE spectrum of the 2$^{nd}$ QD, used for CPT experiments, is displayed in Fig. S2, with the emission separated into the two $X^+_{h0h0}$ emission lines, $\Uparrow^0 - S^0_{+1/2}$ and $\Downarrow^0 - S^0_{-1/2}$. If circularly polarized excitation is used, the lines that emit from $\Uparrow^0 - S^0_{+1/2}$ are excited with $\sigma^-$, and the lines that emit from $\Downarrow^0 - S^0_{-1/2}$ are excited with $\sigma^+$. The spectrum is similar to that of the 1$^{st}$ QD, but the separation between the $T^1_{\pm 1/2}$ and $T^1_{\pm 5/2}$ lines is not as large.

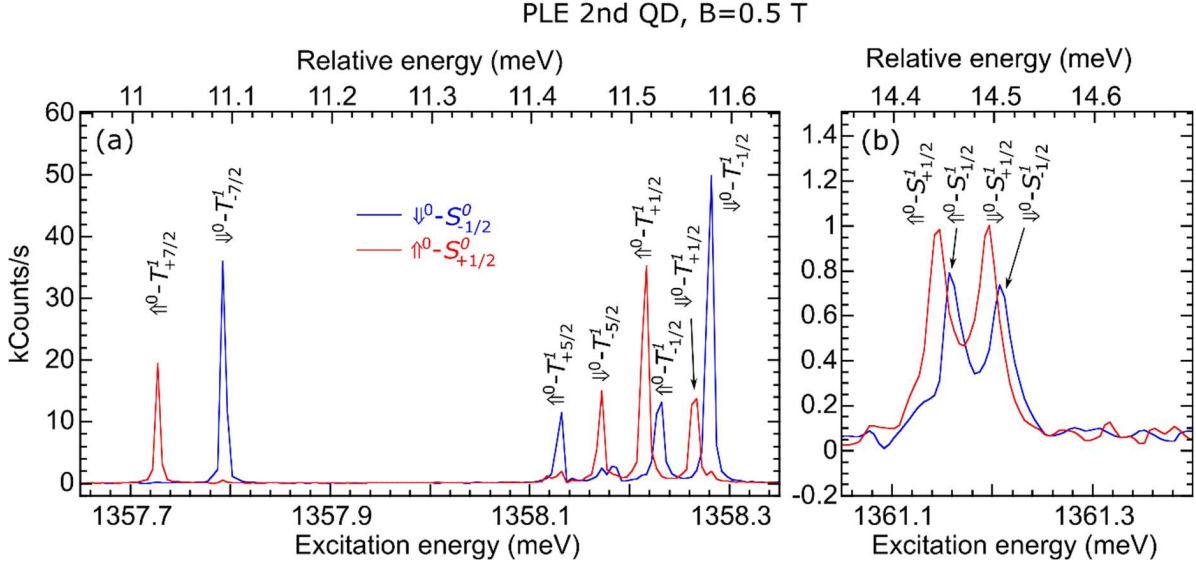

Fig. S2. PLE spectra of (a) the triplet and (b) the singlet transitions for the 2$^{nd}$ QD at B=0.5 T, $V_{bias} = -0.87$ V, with linearly polarized excitation at 0.3 µW.

## III. Triplet pumping times

Figure S3 displays additional information on the pumping rates for the triplet transitions. In Fig. S3(a), PLE spectra are plotted at a bias of $V_{bias} = -0.97$ V, where spin relaxation is slow enough for optical pumping to occur. An additional laser at 1388 meV is on continuously to prevent complete optical pumping, also giving rise to the offset of about 50 kCounts/s. The two spectra are obtained by plotting PLE separately for the $\Uparrow^0 - S^0_{+1/2}$ and $\Downarrow^0 - S^0_{-1/2}$ lines. The Zeeman splittings of the transitions are $(g_{e0} - g_{h0})\mu_B B$ for $S^0_{\pm 1/2}$, $(g_{e0} + g_{h1})\mu_B B$ for $T^1_{\pm 7/2}$, and $(-g_{e0} + g_{h1})\mu_B B$ for $T^1_{\pm 5/2}$. For $T^1_{\pm 1/2}$ and $S^1_{\pm 1/2}$ the four lines are at $(\pm g_{e0} \pm g_{h0})\mu_B B$. From measuring transitions as a function of B, we obtain $g_{h0} = 2.13$ and $g_{h1} = -1.92$. $g_{e0}$ appears to vary between -0.3 and -0.6, presumably due to variable Overhauser shifts from nuclear spin polarization.

As in the main text, the spin pumping times are observed by first pulsing a linearly-polarized laser at 1388 meV for 30 ns to randomize the hole spin state. Trion transitions are quite broad ($> 200$ µeV) at this excitation energy, so it is a good assumption that the two hole spin states are excited equally. Spin flip processes during non-radiative relaxation and emission can then result in roughly equal populations of hole spin orientations. This is followed by a 100 ns pulse resonant with one of the triplet transitions. Figure S3(b) plots the time-correlated-photon-counting trace of emission from $S^0_{-1/2}$ for all of the transitions that predominantly emit from this line, indicated by the colored arrows in Fig. S3(a). At a peak power of 1 µW, the decay time constants are 29.4, 11.2, 7.3, and 20.5 ns for the $\Downarrow^0 - T^1_{-7/2}$, $\Uparrow^0 - T^1_{+5/2}$, $\Uparrow^0 - T^1_{-1/2}$, and $\Downarrow^0 - T^1_{-1/2}$ transitions, respectively. All of the transitions result in optical pumping as a result of the non-radiative relaxation and emission that have some probability of returning the system to the opposite hole spin state. The diagrams in Fig. S3(c-e) illustrate the dominant processes that determine the pumping times and why these vary for the different transitions. The dominant processes are determined from the spectrally resolved PLE in Fig. S3(a) and similar spectra. From inspection of these diagrams it can be seen that the electron spin is preserved during non-radiative relaxation while the excited hole spin is not.

For the $T^1_{\pm 7/2}$ transitions, the excitation, relaxation and emission cycle shown in Fig. S3(c) primarily returns the system to its original hole spin state, giving two cycling transitions. This cycle is not perfect, however, due to some relaxation to the opposite spin state as well as the electron sometimes tunneling out of the QD. This gives a relatively long pumping time of 29.4 ns. For the $T^1_{\pm 5/2}$ transitions, even though they appear to form a double two-level system based on the transition selection rules, relaxation and emission primarily return the system to the *opposite* hole spin state, giving a fast pumping time of 11.2 ns. The $T^1_{\pm 1/2}$ transitions are quite different for the expected vertical transitions and the unexpected diagonal transitions. Excitation of a vertical transition primarily returns the system to the same hole spin state, while excitation of a diagonal transition primarily returns the system to the opposite hole spin state. The difference in these pumping times (20.5 ns vs 7.3 ns) is consistent with this picture. In order to completely account for the pumping rates, one must also consider other relaxation pathways: first, the electron tunneling out of the QD followed by a hole tunneling back to the reservoir; second, emission from $X^+_{h0h1}$ into $h_1$, followed by relaxation to $h_0$; third, emission directly from $X^+_{h0h1}$ to $h_0$. These processes are not very likely in most cases but may still affect the pumping times.

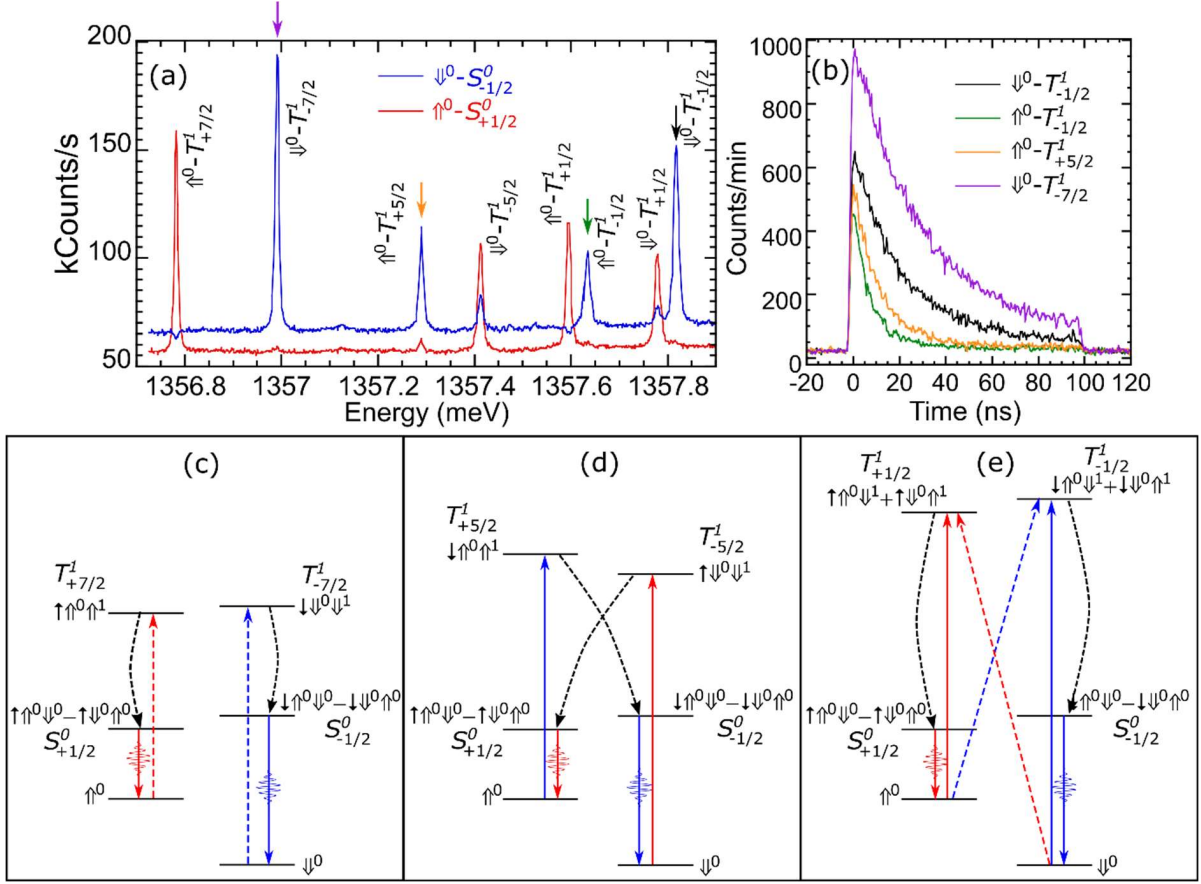

Fig. S3. (a) PLE of the triplet transitions at B=1.5 T and $V_{bias} = -0.97$ V, with 1 µW linearly polarized excitation, separating emission into the $\Uparrow^0 - S^0_{+1/2}$ and $\Downarrow^0 - S^0_{-1/2}$ lines. A non-resonant cw laser at 1388 meV is on at 9 µW to prevent complete optical pumping. Colored arrows indicate the transitions for which the pumping times are measured in (b). (b) Time-correlated-photon-counting of emission from the $\Downarrow^0 - S^0_{-1/2}$ line when driving four different triplet transitions. (c-e) Energy level diagrams showing excitation, non-radiative relaxation, and emission for $T^1_{\pm 7/2}$, $T^1_{\pm 5/2}$, and $T^1_{\pm 1/2}$, with curved, dashed arrows indicating the dominant relaxation path.

## IV. Spin-orbit coupling
### a. Hole spin-orbit coupling interaction (SOC)

In III-V materials there is a built-in spin-orbit coupling in bulk which leads to the structure at the top of the valence band. In addition, the confinement potential leads to spin-orbit coupling between confined states. The latter spin-orbit interaction potential is given by

$$\widehat{\boldsymbol{h}} \cdot \boldsymbol{S} = \gamma_z (\boldsymbol{\mathcal{E}}_\perp \times \widehat{\boldsymbol{p}}_\perp) \cdot \boldsymbol{S}_z + \gamma_\perp (\boldsymbol{\mathcal{E}}_z \times \widehat{\boldsymbol{p}}_\perp) \cdot \boldsymbol{S}_\perp, \qquad (S1)$$

where $\boldsymbol{S}$ stands for the hole pseudospin operator in the Hilbert subspace of heavy hole (HH) states $J_z = \pm 3/2$. Indices $\perp$, $z$ stand for in-plane and vertical (growth) directions. Here we consider the dominant Rashba coupling arising from the effective electric field $\mathcal{E}$ of the nanostructure [1]. The motion in the electric field $\mathcal{E}$ gives rise to an effective magnetic field $\propto \mathcal{E} \times \hat{p}$. The coupling strengths $\gamma_z$ are $\gamma_\perp$ are proportional to effective HH g-factors for magnetic fields in the vertical (Faraday) and in-plane (Voigt) directions, respectively. The term containing $S_z$ on the rhs of Eq. (S1) is responsible for the shift of the triplet states discussed in the main text. It conserves the pseudospin and it involves the motion between $p_x$ and $p_y$ states [2]. The term containing $S_\perp$ determines the nominally forbidden transitions described in the paper. That term is the usual Rashba term arising from asymmetry in the growth direction, and it flips the pseudospin through a precession in the effective Voigt magnetic field.

In the case of electrons [2] $\gamma_z$ and $\gamma_\perp$ are equal to one another, and they occur in the lowest order of $\mathbf{k} \cdot \hat{p}$ perturbation theory between the conduction band and valence band states $|J| = 3/2$. For holes, $\gamma_z$ is obtained in the same order of perturbation theory, whereas $\gamma_\perp$ is given in a higher order of perturbation theory by mixing in LH states [2]. The latter amounts to a projection of the four-dimensional valence band Hilbert space onto the two-dimensional pseudospin subspace $|J_z| = 3/2$ [2]. Below we describe the effect of the interaction potential in Eq. (S1) on a single hole pseudospin in a quantum dot (QD). We show how the combination of $\mathbf{k} \cdot \hat{p}$ and strain couplings contribute to $\gamma_\perp$ to build a significant coupling for $h_1$. We conclude by illustrating the spin-flip mixing of excited positive trion states arising from the second term in Eq. (S1) and by showing the mechanism of the triplet state shift that involves the first term in Eq. (S1).

b. **Spin mixing in the $h_1$ state**

SOC results in spin not being a good quantum number, *i.e.*, the spin is no longer aligned with the vertical axis. This turns on nominally forbidden transitions. The observed transitions can be reproduced by a "tilted" pseudospin in the first excited state $h_1$ due to the second term in Eq. (S1):

$$\Uparrow_t^1 = \alpha |h_1\rangle |\Uparrow\rangle + \sum_{i>1} \beta_i |h_i\rangle |\Downarrow\rangle, \quad (S2)$$

$$\Downarrow_t^1 = -\sum_i \beta_{i>1}^* |h_i\rangle |\Uparrow\rangle + \alpha |h_1\rangle |\Downarrow\rangle.$$

This spin tilt occurs due to SOC mixing with higher HH orbital states. The coupling matrix elements $\beta_i$ between QD spinors result from projections of HH-LH mixing as shown in the next section. Eq. (S2) leads to a small tilt angle $\theta$ given by $\alpha = \cos(\theta/2)$, $|\beta| = \sin(\theta/2)$, $|\beta| = \sqrt{\sum_{i>1} |\beta_i|^2}$. State $h_1$ has a vertical spin projection value of $\langle \Uparrow_t^1 \rangle_z = (3/2)(\alpha^2 - |\beta|^2) = (3/2)\cos\theta$. In the case of vanishing SOC, we have the limit For $\beta \to 0$: $\sin(\theta/2) \to 0$, $\cos(\theta/2) \to 1$, therefore $\theta \to 0$ and $\langle \Uparrow_t^1 \rangle_z \to 3/2$.

We find that the experimental data from the main text can be represented in a simplified model in which the spin-orbit coupling has a strong effect on the $h_1$ pseudospin, producing the spin tilt discussed above, and a relatively small effect on the $h_0$ pseudospin. This comes from relatively smaller admixtures of excited orbitals with the ground state through the second term in Eq. (S1), based on symmetry properties and energy separations. Therefore, we neglect spin tilt for $h_0$.

### c. Heavy hole SOC in QD excited states

In bulk, the HH and LH are coupled through $J_\pm$, $J_\pm^3$ operators [1]. The portion of the interaction potential with the form $r_{41}^{8v,8v}(k_- J_+ - k_+ J_-)$ dominates through its large coefficient $r_{41}^{8v,8v}$ [1], and it couples HH to LH states (e.g., $J_z = 3/2$ to $J_z = 1/2$) and LH to LH states (e.g., $J_z = 1/2$ to $J_z = -1/2$). This amounts to a three-step indirect pseudospin coupling between the HH states ($\Delta J_z = \pm 1$ at each step). The portion of the interaction potential with the form $r_{41}^{8v,8v}(k_- J_+^3 - k_+ J_-^3)$ can in principle couple the HH states directly [1]. Nevertheless, its coefficient $r_{42}^{8v,8v}$ is three orders of magnitude smaller than $r_{41}^{8v,8v}$ and therefore this part is negligible (Chapter 6 in Ref. [1]). Using the basis set $\{|J_z = 3/2\rangle, |J_z = 1/2\rangle, |J_z = -1/2\rangle, |J_z = -3/2\rangle\}$ the effective resulting coupling in bulk is

$$r_{41}^{8v,8v}(k_- J_+ - k_+ J_-) = r_{41}^{8v,8v} E_z \begin{bmatrix} 0 & \frac{i\sqrt{3}}{2} k_- & 0 & 0 \\ -\frac{i\sqrt{3}}{2} k_+ & 0 & ik_- & 0 \\ 0 & -ik_+ & 0 & \frac{i\sqrt{3}}{2} k_- \\ 0 & 0 & -\frac{i\sqrt{3}}{2} k_+ & 0 \end{bmatrix}. \quad (S3)$$

Ref. [1] describes the derivation of coefficient $r_{41}^{8v,8v}$ as a *third order perturbation* coupling between the top of the valence band ($|J| = 3/2$) and the conduction band: two orders in $\mathbf{k} \cdot \hat{\mathbf{p}}$ and one order in the external potential $V$ [1]. The effective coupling between the HH states is obtained with two additional orders in $\mathbf{k} \cdot \hat{\mathbf{p}}$ perturbation theory via higher electron conduction bands [1]. In quantum wells (QWs), the HH isospin Hamiltonian is obtained as a sequence between HH and LH vertically-confined minibands of different parities [1]. In quantum dots, the 2D minibands are replaced by 3D confinement.

In QDs the minibands differ in parity from one another in the lateral coordinates. We find that the $\mathbf{k} \cdot \hat{\mathbf{p}}$ coupling is compounded by the lateral strain ubiquitous in QDs [3–6]. This is a key feature of QDs since it replaces two $\mathbf{k} \cdot \hat{\mathbf{p}}$ coupling terms with higher conduction bands that was necessary in QWs [1]. This is described by Bir-Pikus theory, giving an effective Hamiltonian for $|J| = 3/2$ holes:

$$\begin{bmatrix} P+Q & 0 & R & 0 \\ 0 & P-Q & 0 & R \\ R^* & 0 & P-Q & 0 \\ 0 & R^* & 0 & P+Q \end{bmatrix}, \quad (S4)$$

which couples LH to HH states through $\Delta J_z = \mp 2$ terms on the third diagonals, which can be represented as $R J_+^2 + R^* J_-^2$. Symmetry-based Bir & Pikus theory gives $P, R$ and $Q$:

$$P = a_v \sum_i \epsilon_{ii}, \quad R = id\epsilon_{xy} - \frac{b\sqrt{3}}{2}(\epsilon_{xx} - \epsilon_{yy}), \quad Q = b\left(\frac{\epsilon_{xx} + \epsilon_{yy}}{2} - \epsilon_{zz}\right) \quad (S5)$$

where $a_v, d, b$ are the deformation potentials. Eq. (S5) has been used for QD ground shell holes $h_0$ to describe, *e.g.*, elliptical polarization [2,4–7]. In our case, the term $\epsilon_{xx} - \epsilon_{yy}$ is activated by the anisotropy of $h_1$ orbital and does not require anisotropy of the strain. This gives a strong coupling. The total effective Hamiltonian from Eqs. (S3) and (S4) is:

$$\begin{bmatrix} P+Q & \frac{i\sqrt{3}}{2}r_{41}^{8v,8v}E_z k_- & R & 0 \\ -\frac{i\sqrt{3}}{2}r_{41}^{8v,8v}E_z k_+ & P-Q & ir_{41}^{8v,8v}E_z k_- & R \\ R^* & -i\,r_{41}^{8v,8v}E_z k_+ & P-Q & \frac{i\sqrt{3}}{2}r_{41}^{8v,8v}E_z k_- \\ 0 & R^* & -\frac{i\sqrt{3}}{2}r_{41}^{8v,8v}E_z k_+ & P+Q \end{bmatrix} \quad (S6)$$

In our proof-of principle model for flat QDs the vertical and lateral degrees of freedom are separated. All HH orbital share the same vertical confinement state, which is also the case for LHs. The LH states have a higher zero-point energy compared to HH due to the lower effective mass $m^*_{LH,z} < m^*_{HH,z}$ in the growth direction. In the lateral direction the effective masses are reversed ($m^*_{LH,\perp} > m^*_{HH,\perp}$) which gives smaller spacing between QD LH orbitals compared to HH orbitals. The pseudospin Hamiltonian in Eq. (S1) is obtained from Eq. (S6) by projecting the HH-LH mixing from the four-dimensional Hilbert space above to the two-dimensional HH pseudospin space in Löwdin's second-order perturbation theory [1]. This includes two types of terms (paths) that give the pseudospin-flip couplings $\beta_{i,i>1}$ between QD HH states $h_1$ and $h_{i,i>1}$ in Eq. (S2):

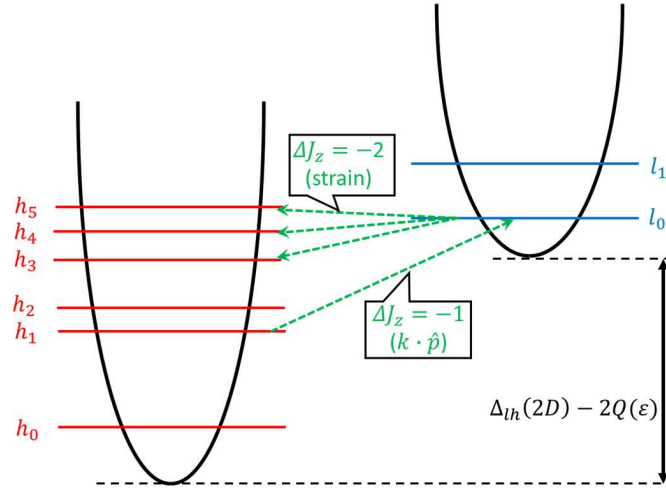

Fig. S4. Path (a) for tilting $h_1$ pseudospin by adding a d-component with opposite spin through a $k \cdot \hat{p}$ coupling proportional to $J_-$, followed by strain coupling proportional to $\epsilon J_-^2$.

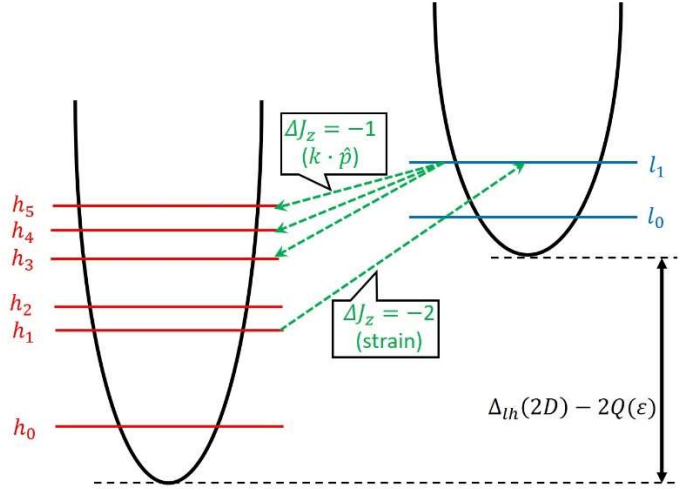

Fig. S5. Path (b) for tilting $h_1$ pseudospin by adding a d-component with opposite spin through strain coupling proportional to $\epsilon J_-^2$ followed by a $k \cdot \hat{p}$ coupling proportional to $J_-$.

- Path (a, Fig. S4) via a $\Delta J_z = -1$ angular momentum-flip $h_1 \to l_0$ due to the $k_+$ coupling in the $J_-$-SOC followed by a $\Delta J_z = -2$ angular momentum-flip $l_0 \to h_3$ through the strain coupling proportional to $J_-^2$.
- Path (b, Fig. S5) via a $\Delta J_z = -2$ angular momentum-flip $h_1 \to l_1$ through the strain coupling proportional to $J_-^2$ followed by a $\Delta J_z = -1$ angular momentum-flip $l_1 \to h_3$ through $k_+$ in the $J_-$-SOC.

In Fig. S4 and Fig. S5 $\Delta_{lh}(2D)$ reflects a shift between lateral confinement potentials for the lowest HH and LH due to the difference in vertical confinement lowest minibands. The additional shift Q reflects the strain coupling in Eq. (S5). Below we denote the energy spacing between HH orbitals by $\Delta\varepsilon_{ij}^{hh} = \varepsilon_i^h - \varepsilon_j^h$ and between HH and LH orbitals by $\Delta\varepsilon_{ij}^{lh} = \varepsilon_i^l - \varepsilon_j^h$. The $s$-like HH state has a negligible contribution to the spin tilts of $h_1$: paths (a) could include in principle a second step $l_0 \to h_0$. Nevertheless, the latter can be ignored: the matrix element of the anisotropic strain between $l_0$ and $h_0$ is smaller due to relatively small anisotropies of these states and the energy separation $|\Delta\varepsilon_{00}^{lh}| > |\Delta\varepsilon_{03}^{lh}|$ reduces further this relative contribution. Similarly, path (b) can be completed in principle with $l_1 \to h_0$. Nevertheless, the large energy difference $|\Delta\varepsilon_{10}^{lh}| > |\Delta\varepsilon_{13}^{lh}|$ reduces significantly the contribution of this path. The summation of the two paths above give the mixing $\beta_3$ of the $d_{xx}$-like orbital $h_3$ with $h_1$:

$$\beta_3 = -\frac{3ir_{41}^{8v,8v} E_z b}{8 \Delta\varepsilon_{13}^{hh}} \left\{ \langle \Uparrow_h^1 | k_- | \Uparrow_l^0 \rangle \langle \Uparrow_l^0 | \epsilon_{xx} - \epsilon_{yy} | \Downarrow_h^3 \rangle \left[ \frac{1}{\Delta\varepsilon_{01}^{lh}} + \frac{1}{\Delta\varepsilon_{30}^{hl}} \right] + \right. \quad (S7.1)$$

$$\langle \Uparrow_h^1 | \epsilon_{xx} - \epsilon_{yy} | \Downarrow_l^1 \rangle \langle \Downarrow_l^1 | k_- | \Downarrow_h^3 \rangle \left[ \frac{1}{\Delta\varepsilon_{11}^{lh}} + \frac{1}{\Delta\varepsilon_{31}^{hl}} \right] +$$

$$\left. \langle \Uparrow_h^1 | \epsilon_{xy} | \Downarrow_l^2 \rangle \langle \Downarrow_l^2 | k_- | \Downarrow_h^3 \rangle \left[ \frac{1}{\Delta\varepsilon_{21}^{lh}} + \frac{1}{\Delta\varepsilon_{32}^{hl}} \right] \right\}$$

The first term in Eq. (S7.1) corresponds to path a, and the last two terms to path b. There is a similar contribution to the spin tilting of $h_1$ from the $d_{yy}$-dominated orbital $h_5$. The latter has the same parity as $h_3$. Since $|\Delta\varepsilon_{15}^{hh}| > |\Delta\varepsilon_{13}^{hh}|$ this results in a smaller mixing $\beta_5$, where

$$\beta_5 = -\frac{3ir_{41}^{8v,8v} E_z b}{8\,\Delta\varepsilon_{15}^{hh}} \Bigg\{ \langle \Uparrow_h^1 | k_- | \Uparrow_l^0 \rangle \langle \Uparrow_l^0 | \epsilon_{xx} - \epsilon_{yy} | \Downarrow_h^5 \rangle \left[\frac{1}{\Delta\varepsilon_{01}^{lh}} + \frac{1}{\Delta\varepsilon_{50}^{hl}}\right] + \quad (S7.2)$$

$$\langle \Uparrow_h^1 | \epsilon_{xx} - \epsilon_{yy} | \Downarrow_l^1 \rangle \langle \Downarrow_l^1 | k_- | \Downarrow_h^5 \rangle \left[\frac{1}{\Delta\varepsilon_{11}^{lh}} + \frac{1}{\Delta\varepsilon_{51}^{hl}}\right] +$$

$$\langle \Uparrow_h^1 | \epsilon_{xy} | \Downarrow_l^2 \rangle \langle \Downarrow_l^2 | k_- | \Downarrow_h^5 \rangle \left[\frac{1}{\Delta\varepsilon_{21}^{lh}} + \frac{1}{\Delta\varepsilon_{52}^{hl}}\right] \Bigg\}$$

Finally, there is a contribution to the spin tilting of $h_1$ from the $d_{xy}$-dominated orbital $h_4$. Nevertheless, the dipole matrix element between the ground state and this orbital is negligible and therefore the admixture of $h_4$ does not contribute significantly to the observed transitions.

Eqs. (S7) implies that $\Uparrow_h^1$ is not a pure HH state: it includes a first-order admixture of LH states. A significant component with $\Uparrow_l^0$ would partially allow the nominally forbidden transitions in Fig. 3(c) but with opposite polarization selection rules. For example, if the triplet state $(\uparrow\Uparrow_h^0\Downarrow_h^1 + \uparrow\Downarrow_h^0\Uparrow_h^1)/\sqrt{2}$ is given a small component $\uparrow\Downarrow_h^0\Uparrow_l^0$ through the second term, the transition from the ground state $\Downarrow_h^0$ is then allowed with $\sigma^+$ polarization, opposite to that observed. An admixture with $\Downarrow_l^1$, however, would not turn on these transitions since the resulting states would still give spin-forbidden transitions and weak overlap between $l_1$ and $e_0$. The observed transitions suggest that the main contribution to this admixture comes from the $\Downarrow_l^1$ LH state and a negligible contribution of $\Uparrow_l^0$. This is consistent with the HH-LH strain coupling dominating over the HH-LH $\mathbf{k}\cdot\hat{\mathbf{p}}$ coupling for these lowest excited states, which are anisotropic and therefore do not require anisotropy of the strain field.

d. **Spin mixing for the $h_1 h_0$ trion**

To illustrate the effect of hole spin-orbit coupling on positive "tilted spin" trions, we choose the states with electron spin state $\uparrow$. We emphasize that the $h_0$ and $e_0$ spins are not tilted:

$$T_{t,+7/2}^1 = \uparrow (\Uparrow_t^1 \Uparrow^0 - \Uparrow^0 \Uparrow_t^1)/\sqrt{2} \quad (S8.1)$$

$$T_{t,-5/2}^1 = \uparrow (\Downarrow_t^1 \Downarrow^0 - \Downarrow^0 \Downarrow_t^1)/\sqrt{2} \quad (S8.2)$$

$$T_{t,+1/2}^1 = \uparrow (\Uparrow_t^1 \Downarrow^0 - \Downarrow^0 \Uparrow_t^1)/2 + \uparrow (\Downarrow_t^1 \Uparrow^0 - \Uparrow^0 \Downarrow_t^1)/2 \quad (S8.3)$$

$$S_{t,+1/2}^1 = \uparrow (\Uparrow_t^1 \Downarrow^0 - \Downarrow^0 \Uparrow_t^1)/2 - \uparrow (\Downarrow_t^1 \Uparrow^0 - \Uparrow^0 \Downarrow_t^1)/2 \quad (S8.4)$$

We use the indices *0* and *t* to denote states before and after turning on the SOC (tilting the $h_1$ pseudospin), respectively. (Note that here the states are explicitly antisymmetric while in the main text abbreviated notation is used that is not.) The states in Eq. (S8) are split by the hole-hole exchange $\Delta_{hh}$. If the electron-hole exchange is the same for $h_0$ and $h_1$ ($\Delta_{eh}^0 =$ $\Delta_{eh}^1$), $T_{t,+1/2}^1$ and $S_{t,+1/2}^1$ are equal admixtures of $\uparrow (\Uparrow_t^1 \Downarrow^0 - \Downarrow^0 \Uparrow_t^1)/\sqrt{2}$ and $\uparrow (\Downarrow_t^1 \Uparrow^0 - \Uparrow^0 \Downarrow_t^1)/\sqrt{2}$. $\Delta_{eh}^0 \neq \Delta_{eh}^1$ changes this admixture and brings additional energy shifts.

Based on the tilted hole spin equation Eq. (S2), the tilted-spin trion states in Eq. (S8) can be expanded in terms of trion states unperturbed by SOC:

$$T_{t,+7/2}^1 = \alpha\, T_{0,+7/2}^1 + \sum_{i>1} (\beta_i/2)\left(T_{0,+1/2}^i - S_{0,+1/2}^i\right) =$$

$$= (\alpha/\sqrt{2})\uparrow |\Uparrow^1\Uparrow^0 - \Uparrow^0\Uparrow^1\rangle + \sum_{i>1}(\beta_i/\sqrt{2})\uparrow |\Downarrow^i\Uparrow^0 - \Uparrow^0\Downarrow^i\rangle, \quad (S9.1)$$

$$T_{-5/2}^1 = \alpha\, T_{0,-5/2}^1 - \sum_{i>1}(\beta_i^*/\sqrt{2})\left(T_{0,+1/2}^i + S_{0,+1/2}^i\right) =$$

$$= (\alpha/\sqrt{2})\uparrow |\Downarrow^1\Downarrow^0 - \Downarrow^0\Downarrow^1\rangle - \sum_i (\beta_i^*/\sqrt{2})\uparrow |\Uparrow^i\Downarrow^0 - \Downarrow^0\Uparrow^i\rangle, \quad (S9.2)$$

$$T_{+1/2}^1 = \alpha\, T_{0,+1/2}^1 + (1/\sqrt{2})\sum_{i>1}\left(\beta_i T_{0,-5/2}^i - \beta_i^* T_{0,+7/2}^i\right) = \quad (S9.3)$$

$$= (\alpha/2)\uparrow (|\Uparrow^1\Downarrow^0 - \Downarrow^0\Uparrow^1\rangle + |\Downarrow^1\Uparrow^0 - \Uparrow^0\Downarrow^1\rangle) + (1/2)\sum_{i>1}\uparrow\left(\beta_i|\Downarrow^i\Downarrow^0 - \Downarrow^0\Downarrow^i\rangle + \beta_i^*\uparrow|\Uparrow^i\Uparrow^0 - \Uparrow^0\Uparrow^i\rangle\right),$$

$$S_{+1/2}^1 = \alpha\, S_{0,+1/2}^1 + (1/\sqrt{2})\sum_{i>1}\left(\beta_i T_{0,-5/2}^i - \beta_i^* T_{0,+7/2}^i\right) = \quad (S9.4)$$

$$= (\alpha/2)\uparrow (|\Uparrow^1\Downarrow^0 - \Downarrow^0\Uparrow^1\rangle - |\Downarrow^1\Uparrow^0 - \Uparrow^0\Downarrow^1\rangle) + (1/2)\sum_{i>1}\uparrow\left(\beta_i|\Downarrow^i\Downarrow^0 - \Downarrow^0\Downarrow^i\rangle - \beta_i^*\uparrow|\Uparrow^i\Uparrow^0 - \Uparrow^0\Uparrow^i\rangle\right).$$

These expressions show the small contributions of the spin-flipped hole via excited states to the nominally forbidden optical transitions involving the $e_0 h_1 h_0$ trion. The mixing of triplet and singlet states is a signature of asymmetric exchange.

e. **Shift of triplet states in $h_1 h_0$ and $h_2 h_0$ trions**

The shift of the triplet states can be understood by considering the first term in Eq. (S1) for a single hole and deriving the effect of two such terms for the for two holes $i=1,2$ in the trion states.

One can start with the single-particle coupling

$$h_z^{(i)} S_z = \gamma_z \left(\mathcal{E}_\perp \times \hat{p}_\perp^{(i)}\right) \cdot S_z^{(i)} = -i\hbar\gamma_z\left[\partial_{x_i} V(\vec{r}_i)\partial_{y_i} - \partial_{y_i} V(\vec{r}_i)\partial_{x_i}\right] \quad (S10)$$

Since the potential $V(\vec{r}_i)$ has a dominant reflection-symmetric character, the two derivatives require two orbitals that differ in parity in both $x$ and $y$ directions. For the lowest excited orbital $h_1$, which is $p_x$-like, this is activated by coupling to the next excited orbital $h_2$, which is $p_y$-like. Then one can derive the effect of the coupling between two-hole states given by the sum of single-particle couplings:

$$H_{SOC}^z = h_z^{(1)} S_z^{(1)} + h_z^{(2)} S_z^{(2)} \quad (S11)$$

This couples states with the same spin. For the case with parallel hole spins, *e.g.*, trions $T^1_{0,\pm 7/2}$ and $T^2_{0,\pm 7/2}$, it pushes the former downward and the latter upward. It also couples trion states in which the holes have opposite hole spins, *i.e.*, triplet $T^1_{0,\pm 1/2}$ to singlet states $S^2_{0,\pm 1/2}$ (asymmetric exchange). The level repulsion between the latter is smaller since these are separated by a larger energy. This can be shown as follows:

$$\langle T^1_{0,\pm 7/2}|H^z_{SOC}|T^2_{0,\pm 7/2}\rangle = \langle T^1_{0,\pm 5/2}|H^z_{SOC}|T^2_{0,\pm 5/2}\rangle = \pm h^{yx}_z \qquad (S12)$$

where $h^{yx}_z = h^{yx(i)}_z = \langle h_1|h^{(i)}_z|h_2\rangle$ is the orbital single-particle matrix element for $i = 1,2$ in Eq. (S11).

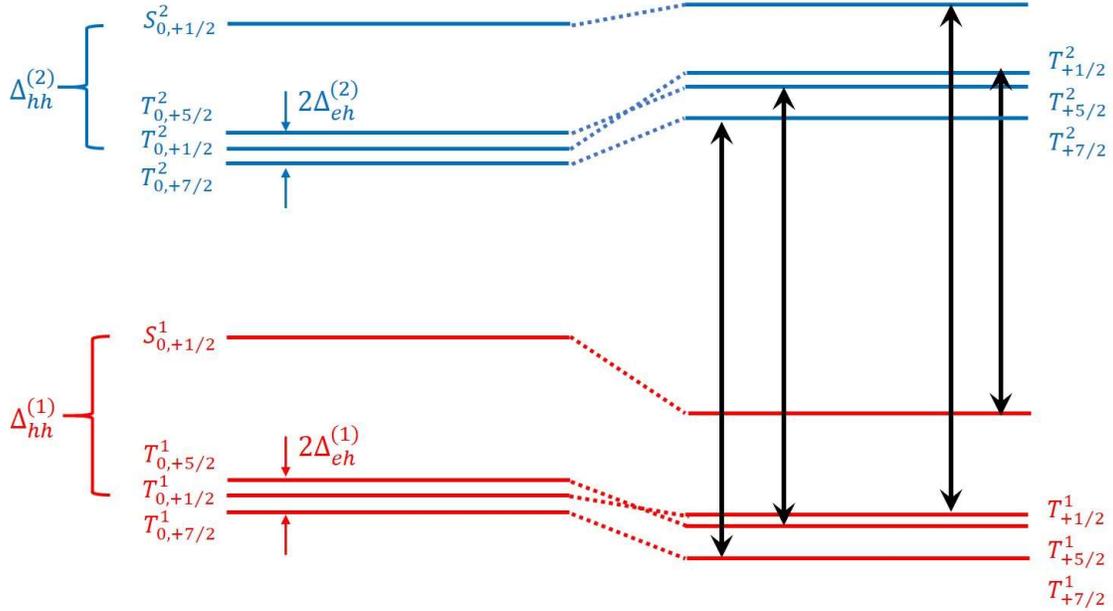

Fig. S6. Schematic level repulsion between the two lowest trion shells. The double arrows show the levels coupled by the interaction in Eq. (S11). The derivation of energy shifts below ignores the small e-h exchange splitting $\Delta^{(1)}_{eh}$ and $\Delta^{(2)}_{eh}$.

For illustration purposes, we can use an approximation that neglects the e-h exchange and Zeeman terms: $E_{T^2_{0,\pm 7/2}} = E_{T^2_{0,\pm 1/2}} = E_{T^2_{0,\pm 5/2}} = E^{(2)}_0 - \Delta^{(2)}_{hh}/2$, and $E_{S^2_{0,\pm 1/2}} = E^{(2)}_0 + \Delta^{(2)}_{hh}/2$. Here, we use $\Delta^{(1)}_{hh}$ and $\Delta^{(2)}_{hh}$ to label the h-h exchange in the two shells. With this, the first order perturbation approximation gives a downward shift of $T^1_{0,\pm 7/2}$ and an upward shift of $T^2_{0,\pm 7/2}$ illustrated in Fig. S6:

$$\delta E_{T^1_{0,\pm 7/2}} \cong \frac{|h^{yx}_z|^2}{E_{T^1_{0,\pm 7/2}} - E_{T^2_{0,\pm 7/2}}} \cong \frac{|h^{yx}_z|^2}{E^{(1)}_0 - E^{(2)}_0 - (\Delta^{(1)}_{hh} - \Delta^{(2)}_{hh})/2} < 0, \ \delta E_{T^2_{0,\pm 7/2}} = -\delta E_{T^1_{0,\pm 7/2}} > 0 \qquad (S13)$$

For trion states with opposite hole spins, the coupling matrix element given by Eq. (S11) amounts to:

$$\langle T^1_{0,\pm 1/2} | H^z_{SOC} | S^2_{0,\pm 1/2} \rangle = -h^{yx}_z,$$
$$\langle T^2_{\pm 1/2} | H^z_{SO,S} | S^1_{\pm 1/2} \rangle = -h^{yx}_z, \qquad (S14)$$
$$\delta E_{T^1_{+1/2}} \cong \frac{|h^{yx}_z|^2}{E_{T^1_{0,+1/2}} - E_{S^2_{0,+1/2}}} \cong \frac{|h^{yx}_z|^2}{E^{(1)}_0 - E^{(2)}_0 - (\Delta^{(1)}_{hh} + \Delta^{(2)}_{hh})/2} < 0$$

Eqs. (S13) and (S14) show that the origin of the different shifts of the two triplets in the $h_1$ sector is the exchange splitting $\Delta^{(2)}_{hh}$:

$$\delta E_{T^1_{0,+1/2}} - \delta E_{T^1_{0,+7/2}} \cong |h^{yx}_z|^2 \frac{\Delta^{(2)}_{hh}}{\left(E^{(1)}_0 - E^{(2)}_0 - \Delta^{(1)}_{hh}/2\right)^2 - \left(\Delta^{(2)}_{hh}/2\right)^2} > 0 \qquad (S15)$$

which means that $T^1_{0,+1/2}$ shifts downward less than $T^1_{0+7/2}$ since $T^1_{0,+1/2}$ is pushed by a higher-energy state, *i.e.*, $S^2_{0,+1/2}$. This is shown schematically in Fig. S6. This inequality holds since in our case the experiment $E^{(2)}_0 - E^{(1)}_0 > \left(\Delta^{(2)}_{hh} - \Delta^{(1)}_{hh}\right)/2$. A similar effect occurs with the $h_2$ shell triplets:

$$\delta E_{T^2_{0,+1/2}} \cong \frac{|h^{xy}_z|^2}{E_{T^2_{0,+1/2}} - E_{S^1_{0+1/2}}} \cong \frac{|h^{xy}_z|^2}{E^{(2)}_0 - E^{(1)}_0 - (\Delta^{(2)}_{hh} + \Delta^{(1)}_{hh})/2} \qquad (S16)$$
$$\delta E_{T^2_{0,+1/2}} - \delta E_{T^2_{0,+7/2}} \cong |h^{xy}_z|^2 \frac{\Delta_1}{\left(E^{(2)}_0 - E^{(1)}_0 - \Delta^{(2)}_{hh}/2\right)^2 - \left(\Delta^{(1)}_{hh}/2\right)^2}$$

which means that $T^2_{0,+1/2}$ shifts upward more than $T^2_{0,+7/2}$ since in our experiment the $h_2$ trion shell is entirely above the $h_1$ trion shell $(E_{S^1_0} < E_{T^2_0.})$.

## V. Determination of the Rabi frequency

The Rabi frequency for Coherent Population trapping is determined by driving the $\Downarrow_0 - T^1_{-1/2}$ transition and measuring the Autler-Townes splitting of the $\Downarrow^0 - S^0_{-1/2}$ emission. The spectra are displayed in Fig. S7(a) for several drive powers, measured with a temperature-tuned, fiber coupled Fabry-Perot interferometer with 2.6 µeV resolution. As illustrated in Fig. S7(b), both the $T^1_{-1/2}$ and $\Downarrow$ states are dressed by the drive field, resulting in Rabi splittings. The splitting is observed in emission from $\Downarrow^0 - S^0_{-1/2}$ at the highest drive power of 16.7 µW, giving $\hbar\Omega = 4.8$ µeV. At the lower powers in Fig. 5(b), the Rabi frequencies are determined by the fact that $\Omega \propto \sqrt{\text{power}}$.

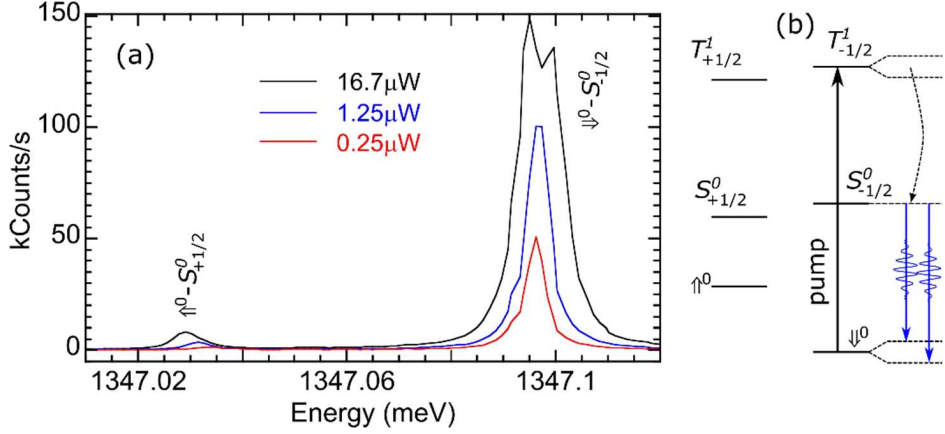

Fig. S7. (a) PL spectra of the 2$^{nd}$ QD for three drive powers while driving the $\Downarrow^0 - T^1_{-1/2}$ transition at $B = 0.4$ T and $V_{bias} = -0.87$ V. (b) Energy level diagram illustrating Autler-Townes splitting, with dashed lines indicating the dressed states.

This measurement was also performed on the 1$^{st}$ QD and compared to measurements of the Mollow triplet when driving the $\Downarrow^0 - S^0_{-1/2}$ transition. From this comparison, the dipole moment is about 19 times smaller for the $\Downarrow^0 - T^1_{-1/2}$ transitions. This ignores the influence of the cavity resonance, which is resonant with $\Downarrow^0 - S^0_{-1/2}$ and detuned from $\Downarrow^0 - T^1_{-1/2}$ by about 0.7 cavity linewidths. We estimate an order of magnitude difference in the dipole moments.

## VI. Coherent Population Trapping Model

Coherent population trapping (CPT) is modeled using the Quantum Optics Toolbox in PYTHON (QUTIP) [8,9]. The double $\Lambda$ system of $T^1_{\pm1/2}$ is approximated as a three level system. This is a reasonable approximation since using $\sigma^+$ drive fields will only couple to the $T^1_{-1/2}$ excited state. Relaxation from $T^1_{-1/2}$ to $S^0_{\pm1/2}$, followed by emission into the ground states is simplified to one step relaxation from $T^1_{-1/2}$ to $\Uparrow^0$ and $\Downarrow^0$. The Hamiltonian is taken from Eqs. 6.45 and 6.46 of Ref. [10]

$$H = \hbar\Delta_1|1\rangle\langle1| + \hbar\Delta_2|2\rangle\langle2| + \tfrac{1}{2}\hbar\Omega_1(\sigma_1 + \sigma_1^\dagger) + \tfrac{1}{2}\hbar\Omega_2(\sigma_2 + \sigma_2^\dagger), \quad (S17)$$

where $|1\rangle$, $|2\rangle$, $|3\rangle$ correspond to $\Downarrow^0$, $\Uparrow^0$, and $T^1_{-1/2}$. $\Delta_1$ and $\Delta_2$ are detunings, $\Omega_1$ and $\Omega_2$ are Rabi frequencies, and $\sigma_i = |i\rangle\langle3|$. This Hamiltonian is written in a rotating frame using the rotating wave approximation. It also assumes that $\omega_1$ only couples to the $|1\rangle \rightarrow |3\rangle$ transition, and $\omega_2$ only couples to the $|2\rangle \rightarrow |3\rangle$ transition. This approximation is reasonable since the spin splitting of 9.8 GHz is much greater than the linewidths of 2 GHz. The off resonant sideband from phase modulation of the laser is also ignored.

Relaxation and dephasing are included with the collapse operators $C_1 = \sqrt{\Gamma}\sigma_1$, $C_2 = \sqrt{\Gamma}\sigma_2$, $C_3 = \sqrt{\gamma_1}|1\rangle\langle 2|$, $C_4 = \sqrt{\gamma_1}|2\rangle\langle 1|$, and $C_5 = \sqrt{\gamma_2}(|1\rangle\langle 1| - |2\rangle\langle 2|)$, where $\Gamma$ is relaxation from the excited state to $|1\rangle$ and $|2\rangle$, $\gamma_1$ is the spin relaxation rate, and $\gamma_2$ is the pure spin dephasing rate. Relaxation is assumed to be equal into $|1\rangle$ and $|2\rangle$ for simplicity even though this is not accurate. Varying the relative relaxation rates does not appear to make a significant difference to the shape of the modelled CPT curves. The steady state value of the density matrix $\rho$ is calculated for a series of probe detunings $\Delta_2$, with $\rho_{33}$ proportional to the emission rate.

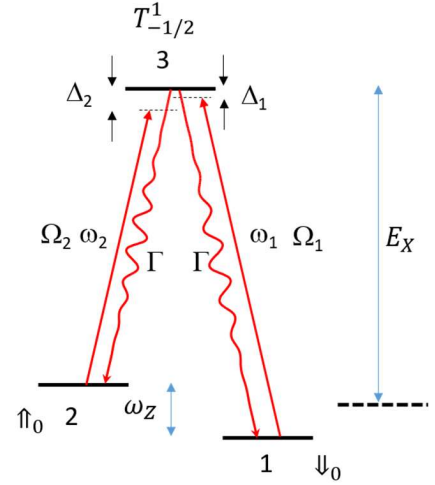

Fig. S8. Energy level diagram of the three level CPT model.

Using this model alone does not fit the experimental data very well. At the experimental Rabi frequencies used, the width of the CPT dip and the wider absorption profile are too narrow compared to experiment. This is due to fluctuations in the Zeeman splitting $\omega_Z$ and in the trion energy $E_X$ that broaden the CPT features. We account for these fluctuations by performing a weighted average over distributions in $\omega_Z$ and $E_X$, with $\omega_Z = \omega_{Z0} + dZ$, $E_X = E_{X0} + dE$ and Gaussian weighting functions $w_Z(dZ) = A_Z e^{-dZ^2/a_Z^2}$, $w_E(dE) = A_E e^{-dE^2/a_E^2}$. The widths of the distributions are $a_Z$ and $a_E$ with normalization constants $A_Z$ and $A_E$. The CPT signal is then given by the following sum over discrete values of $dZ$ and $dE$

$$A \sum_k \sum_i w_Z(dZ_i) w_E(dE_k) \rho_{33}[\Gamma, \gamma_1, \gamma_2, \Omega_1, \Omega_2](\Delta_1, \Delta_2) \tag{S17}$$

The parameters used to calculate $\rho_{33}$ included in brackets do not depend on $dZ_i$ and $dE_k$, while those in parentheses do.

$$\Delta_1 = \omega_1 - \left(E_{X0} + dE_k + \tfrac{1}{2}\omega_{Z0} + \tfrac{1}{2}dZ_i\right) \tag{S18}$$

$$\Delta_2 = \omega_2 - \left(E_{X0} + dE_k - \tfrac{1}{2}\omega_{Z0} - \tfrac{1}{2}dZ_i\right) \tag{S19}$$

The model data in Fig. 5(d) was obtained by first fitting the experimental data at 110 nW, with parameters $A$ (overall scaling factor), $\Gamma$, $a_Z$, $a_E$, $\omega_1$, and $\omega_{Z0}$. The values of $\gamma_1$ and $\gamma_2$ were both fixed at a relatively low value of $5 \times 10^{-4}$ ns$^{-1}$ because the inclusion of significant broadening in $dZ$ made the CPT signal insensitive to spin relaxation and pure dephasing. With this broadening, the condition for observing a strong CPT dip is then $\Omega_1^2/\Gamma \gg 1/T_2^*$. The value of $\Omega_1/2\pi$ was fixed to be 94 MHz from the Autler-Townes splitting measurement, with $\Omega_2 = \Omega_1/4$. This gave a good fit with $A = 1.452 \times 10^6$ counts, $\Gamma = 0.5$/ns, $a_Z/2\pi = 0.0367$ GHz, $a_E/2\pi = 2.02$ GHz, $\omega_1/2\pi = 0.1$ GHz (relative to transition 1), and $\omega_{Z0}/2\pi = 9.805$ GHz. For the CPT data at other powers, $A$, $\Gamma$, $a_Z$, and $a_E$ were kept fixed, with different values of $\Omega_1$ and $\Omega_2$. $\omega_1$ and $\omega_{Z0}$ were free parameters in these fits since the pump detuning and $\omega_{Z0}$ appear to shift slightly with power, presumably due to nuclear polarization effects.

Gaussian fluctuations in the Zeeman splitting give rise to Gaussian decay of spin coherence. The $1/e$ decay time is $T_2^* = 2/a_Z = 8.7$ ns. As discussed in the main text, this is a relatively short value for hole spins. Given the evidence of nuclear polarization effects in the higher power CPT data, it seems possible that the dip width is artificially broadened even at lower powers. Time resolved Ramsey fringe measurements should clarify the value of $T_2^*$.